# Deep Learning Techniques to Make Gravitational Wave Detections from Weak Time-series Data

Yash Chauhan[1*]

[1]*The International School of Bangalore (TISB), Bangalore, India.*

## ABSTRACT

Gravitational waves are ripples in the space time fabric when high *energy* events such as black hole mergers or neutron star collisions take place. The first Gravitational Wave (GW) detection (GW150914) was made by the Laser Interferometer Gravitational-wave Observatory (LIGO) and Virgo Collaboration on September 14, 2015. Furthermore, the proof of the existence of GWs had countless implications from Stellar Evolution to General Relativity. Gravitational waves detection requires multiple filters and the filtered data has to be studied intensively to come to conclusions on whether the data is a just a glitch or an actual gravitational wave detection. However, with the use of *Deep Learning* the process is simplified heavily, as it reduces the level of filtering greatly, and the output is more definitive, even though the model produces a probabilistic result. Our technique, *Deep Learning*, utilizes a different implementation of a one-dimensional convolutional neural network (CNN). The model is trained by a composite of real LIGO noise, and injections of GW waveform templates. The CNN effectively uses classification to differentiate weak GW time series from non-gaussian noise from glitches in the LIGO data stream. In addition, we are the first study to utilize fine-tuning as a means to train the model with a second pass of data, while maintaining all the learned features from the initial training iteration. This enables our model to have a sensitivity of

_________________________________________________________________

*\*Corresponding author: E-mail: yash.chauhan@gmail.com;*

> 100%, higher than all prior studies in this field, when making real-time detections of GWs at an extremely low Signal-to-noise ratios (SNR), while still being less computationally expensive. This sensitivity, in part, is also achieved through the use of deep signal manifolds from both the Hanford and Livingston detectors, which enable the neural network to be responsive to false positives.

## 1. INTRODUCTION

Detection of gravitational waves (GWs) and the study of them can provide a better understanding of the universe from future cosmological observations to testing verifying the reliability of Einstein's General Relativity [1]. The LIGO Scientific Collaboration had been working for close to half-a-century before they were able to make their ground-breaking study in 2015, so there have been multiple studies that explore different, and more improved, forms of GW data analysis. In fact, before the introduction of Deep Learning, there were multiple other methods of searching for GWs, including hierarchical algorithmic search pipelines [2], extensive filtration approaches [3], and matched filtering [4]. However, in addition to requiring extensive manual intervention, most approaches before GW150914 were completely predicated on the use of waveform emulation, which, although quite comprehensive in taking into account the parameters of an event, cannot replicate the signal from an actual event [5-7].

But, as of this moment, the advancements in field Deep Learning have made it possible to revolutionize many aspects of LIGO's data analysis, from glitch classification [8] to parameter inference of BBH events [9]. The most pioneering breakthrough in this field, however, was a study at University of Illinois at Urbana-Champaign [10,11] that effectively used Deep Learning to make real-time GW detections; however, the architecture of their neural network (NN) lacked the complexity to make detections at a sensitivity higher than matched filtering, which is the benchmark for any binary classification study in GWs [12]. Moreover, in this study, even with an extremely high SNR, the sensitivity of the model was still lower than if they had used matched filtering, partly because of the fact that the model was trained using shallow waveform templates that disregarded essential parameters, such as the spin of the BBHs [13-16].

Additionally, our model utilizes LIGO data from both the Hanford and Livingston detectors, providing a higher level of confidence detection than a single detector that was used in the previous study [17]. The primary hypothesis of our study was that our NN would successfully be able to detect gravitational waves from weak signals in a comparable level of accuracy to matched filtering. Currently, matched filtering is most commonly used to make gravitational wave detections with multiple filters and band-passes, which is very computationally extensive, and there is a substantial amount of human intervention required to make definitive detections [4]. Therefore, in the future, if a Deep Learning approach to search pipeline is implemented autonomously, the confidence in making a detection would increase along with a reduction in human error, whilst still using less computational power.

## 2. METHODS

### 2.1 Data Collection

The first step we took to creating our model, effectively classifying gravitational waves (GWs), was data collection. The majority of our data that was loaded onto the cloud was from the Gravitational Wave Open Science Center (GWOSC). Moreover, from this public source we uploaded the time-series data from both the LIGO Hanford and LIGO Livingston detectors for the first five binary black hole (BBH) events. These detectors were the most prominent systems in the initial detection of Gravitational waves, and they continue to provide continuous strain over time data during each observing run. These datasets, after being uploaded to Google Drive, were then loaded onto the Google Collaboratory (Colab) Notebook. Next, we regularized the time series data to be 16384 Hertz per second; this sampling rate was found to be the optimum rate at which to train our CNN model. The rationale behind utilizing this sample rate is that our pre-constructed signals manifolds were created at the sample rate of 16384 Hertz per second, so to avoid any further perturbations to our signal we regularized the signal to 16384 Hertz per second.

## 2.2 Data Manipulation

### 2.2.1 Splitting the input vector to isolate gravitational wave from the noise

The real LIGO event datasets were primarily still made up of noise and glitches; therefore, it was essential to separate the time segment containing the merger from the rest of the noise. Moreover, we created a function that inputted both the data vector and the sampling rate, where the vector would be split in half and only the two central most seconds (where the actual event is located) would be extracted. The rationale behind taking only the middle two seconds was that all the events were in the range of 0.8 - 1.7s, so the reliability of the extraction was extremely high. Additionally, both the actual event, and the rest of the data, were stored in separate files. We used the extracted waveforms from those 5 events (both LIGO Hanford and Livingston data) for testing our *CNN*.

### 2.2.2 Sampling real LIGO noise

To create a robust *Deep learning* model, there needs to be sufficient training data; however, the number of detected events by LIGO were far below the desired amount. Therefore, we had to create our own datasets using the limited amount of data available to us [18,19]. We used the noise files created while splitting the input to vector, to create real LIGO noise where the waveform templates would be injected into, effectively representing continuous LIGO data streams. This gaussian noise was created by randomly sampling the LIGO event data (void of the Gravitational Wave). We created a total of 10,000 noise samples, which included multiple non-gaussian glitches embedded deep in the noise. This enabled us to differentiate between false positives in the search pipeline and actual Gravitational Wave signal, during the testing phase of our model.

## 2.3 Waveform Generator

To optimize the training of our model we had to use 10,000 labeled datasets; however, to achieve this we had to simulate our own waveforms, as the number of events, from the two observing runs, is not of a sufficient number to train the *CNN*. We created our waveforms to delineate the entire progression of the gravitational wave event (BBH coalescence), from the inspiral to the ringdown. The mass parameter of these black holes was randomly sampled between $5M_\odot$ and $75M_\odot$. Furthermore, the rest of the waveform parameters were randomly selected between the bounds of GW1709817 and GW1709818, two polarizing events in terms of magnitude of the event [20,21]. We created the simulated GWs for both the LIGO Hanford and Livingston detectors, which is an advancement that no other prior research has inculcated. This aids in accounting for false positives, as the model compares the signals of both interferometers before making a prediction.

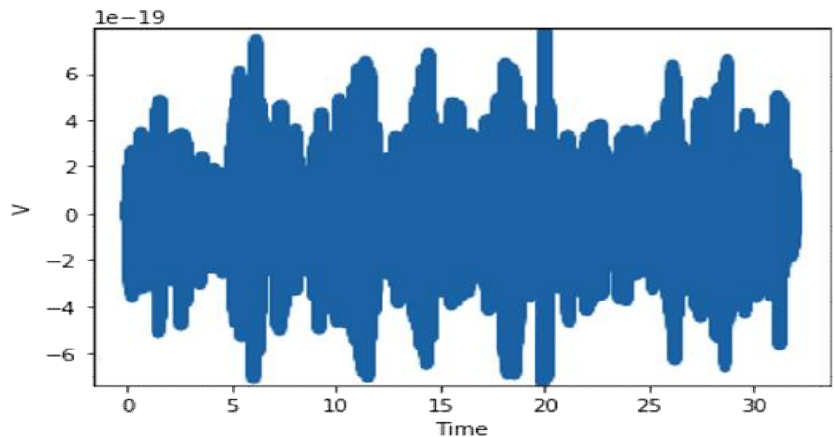

**Fig. 1. Artificially created perturbations from LIGO detectors**
*The plot above is a visual representation of one of the randomly sampled noise files. The scale of the noise would be later increased to an end result of a SNR of 4 when combined with the waveform templates. With an SNR of 4, the noise in comparison to signal is even higher than that of real LIGO data. Therefore, this optimizes the deep filtering model to make real-time detections of Gravitational Waves*

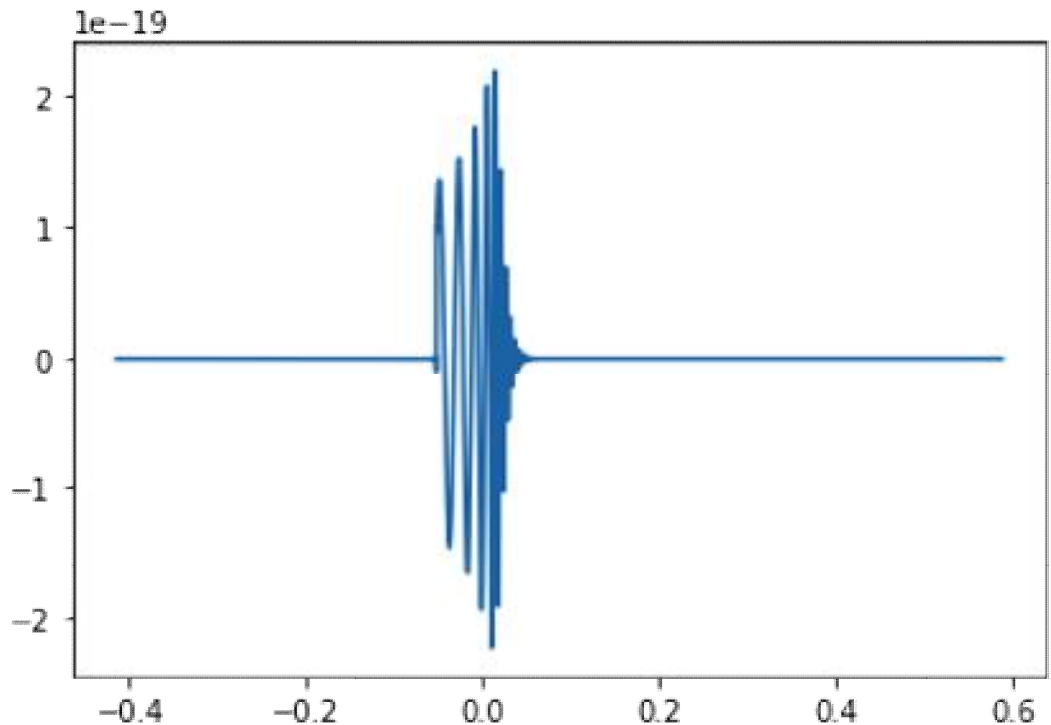

**Fig. 2. Baseline waveform template**
*The y-axis is strain( $1/\sqrt{Hz}$), and the x-axis is time(s). The plot above shows one of the waveform templates we had created that acted as a baseline for us to create our deep signal manifolds. The peak of the event is listed 0 seconds, and you can faintly see the inspiral of both the BBHs before the actual event. Furthermore, the strain of both the real interferometer data and the waveforms are in the same scale, reducing the number of transformations required before the injection into the gaussian noise and glitches*

Additionally, for 25% of the waveforms we changed the time-location of the peaks by an offset of 0.6 seconds, making the *Deep Learning* model more proficient in dealing with time-translations. With this basic signal waveform, we were able to process it until it became indistinguishable with the actual event signal data. This was one the most important transformations in the study because it had a direct effect on whether the model would be able to classify false-positive cases as glitches.

### 2.4 Injection of Waveforms into LIGO Noise

In order to create the training datasets, as discussed above, we had to refactor and scale the waveforms appropriately before injecting the templates into the interferometric noise. The waveform templates from both the LIGO Hanford and Livingston were merged (signal was averaged with the perturbations and locative based discrepancies taken into account) to create a better estimate of a realistic GW signal.

To create the final datasets, we randomly sampled waveform templates and merged them with randomly sampled noise files. In addition, we made us multiple levels of offsets and biases while merging both the waveforms and the gaussian noise, as we required the GW signal to be completely embedded into the noise. We also scaled down the simulated signal to a third of its original strength whilst normalizing the sizes of the signal as well as the noise. In contrast to other studies, the entirety of our created datasets had extremely low SNR (1.5 - 2.5), as when the model can make GW detections with a very weak signal, it can seamlessly be implemented into the LIGO search pipeline which operates at a much higher SNR.

### 2.5 CNN Architecture

Deep Learning models have revolutionized the field of ML, as they are able to classify the faintest of GW signals embedded in the noisiest of the time-domain signals. The supervised learning model we used was a *Convolutional Neural Network (CNN)*, as it is the most efficient and optimized to operate on time-series data [22,23]. The first step when building the model was to assign labels to the training data, regarding whether it contains a GW or not. All the datasets where the gravitational wave was extracted from, which consisted of both gaussian and non-gaussian transient noise (glitches) were given the label [0]. Next, the 5,000 datasets which contained a gravitational wave were given the label [1]. Although, there were 10,000 total datafiles to train the *Neural Network*; we started off by using only 2,000 to optimize training time. Furthermore, we then split those 2000 datasets into 1600 training examples and 400 test cases. From there we reshaped the data to the format (2000 datasets, 16384 samples per second, 1 dimension) to input into the *CNN*.

We utilized a one-dimensional sequential *CNN* model, where the inputted data is forward-propagated through the multiple layers of the model before it classifies the presence of a GW.

The final layer uses the SoftMax activation function to make a prediction on if there is a GW embedded in the time-series data or not. The *CNN* uses 4 convolutional layers with filter sizes of 64, 128, 256 and 512, respectively. Moreover, the kernel sizes used for each of the convolutional layers are 32, 64, 64, and 128, respectively. In addition to this, the standard activation function between convolutional layers is the ReLU function, which essentially works similar to the modulus function. We also used a kernel size of 4 for all the MaxPooling layers. Please take a look at the condensed NetChain figure (Fig. 3) representing the classifier.

The size of the actual model is small, at around 30 MB; however, it encodes a large amount of training examples. Prior studies have required large amounts of computational power and time, to try to duplicate our results. We are the first study to be able to run the entire model locally by using the GPU provided by Google for all Colab Research notebooks.

```
NetChain (Predictor)
      Input              vector (size: 16384)
1     Reshape            matrix (size: 1 x 16384)
2     Convolution        matrix (size: 64 x 16352)
3     Pooling            matrix (size: 64 x 4088)
4     ReLU               matrix (size: 64 x 4088)
5     Convolution        matrix (size: 128 x 4024)
6     Pooling            matrix (size: 128 x 1006)
7     ReLU               matrix (size: 128 x 1006)
8     Convolution        matrix (size: 256 x 942)
9     Pooling            matrix (size: 256 x 236)
10    ReLU               matrix (size: 256 x 236)
11    Convolutional      matrix (size: 512 x 108)
12    Pooling            matrix (size: 512 x 27)
13    ReLU               matrix (size: 512 x 27)
15    Flatten            vector (size: 13824)
16    Dense (ReLU)       vector (size: 128)
17    Dense (ReLU)       vector (size: 64)
18    Dropout            vector (size: 16)
14    Dense (softmax)    vector (size: 2)
      Output             vector (size: 2)
```

**Fig. 3. NetChain representation**

*The figure above illustrates the working of the CNN classifier. The model inputs a time-series vector of 16384 samples in one second, and the size of the vector/matrix is processed as shown by the figure above; however, when it reaches the last layer, the result (Presence of GW) is outputted. This model can easily be implemented for continuous data-streams, as the model can cross-verify between detectors if there is a GW present at that point in time*

After designing the architecture of the CNN and conducting the first training pass (1600 data samples), we fine-tuned our model by freezing the Softmax layer and initializing new nodes in its place [24-26]. Next, we froze the entire *CNN* before the new node to ensure none of the features would be lost through backpropagation. Furthermore, we then used the entirety of the artificially created datasets to fine-tune the fully connected (FC) layers, after which we unfroze the rest of the NN to initiate a second wave of training. The training data used for fine-tuning was the remaining 8400 datasets, a composite of glitches and signal templates. This particular model had a testing accuracy of 99.3% at an SNR of 2, which is considerably higher than all prior deep learning techniques in this field, as the use of transfer learning has never before been used for the classification of GWs, and -it is significant step forward in the progression of *deep learning* in the search pipeline.

## 3. RESULTS

The model produced results that exceeded our expectations, as the sensitivity of the model was even higher than if we had used matched filtering at the same SNR. The model is able to differentiate non-gaussian glitches with real GW signal, as it was trained with examples where the signal was scaled down to a third of its strength. The model only outputs a vector of size 2 that predicts whether there is a gravitational wave present in the time-series; however, it does not output the masses of the Binary Black Holes (BBH) involved.

Our model is better suited to be implemented in the actual LIGO search pipeline in comparison to prior studies [27], because it uses waveforms from both the LIGO Hanford and Livingston detectors to train the CNN. This ensures that when the model is making real-time detections it will be able to compare the time-series data from multiple detectors before making a definite prediction. Thus, minimizing the false detection rate, as the model attaches the label [0] to all glitches from multiple detectors.

The accuracy of this model was at 100% at SNR of 4, where no prior study could achieve this accuracy at a SNR this low. To elaborate further, the *CNN* could classify each and every testing dataset correctly. Please refer to Fig. 4 below this subsection to conceptualize the training and validation metrics of our model.

```
Train on 1600 samples, validate on 400 samples
Epoch 1/12
1600/1600 [==============================] - 1344s 840ms/step - loss: 0.2138 - acc: 0.9200 - val_loss: 1.1921e-07 - val_acc: 1.0000
Epoch 2/12
1600/1600 [==============================] - 1330s 831ms/step - loss: 5.1385e-05 - acc: 1.0000 - val_loss: 1.1921e-07 - val_acc: 1.0000
Epoch 3/12
1600/1600 [==============================] - 1347s 842ms/step - loss: 8.0474e-05 - acc: 1.0000 - val_loss: 1.1921e-07 - val_acc: 1.0000
Epoch 4/12
1600/1600 [==============================] - 1319s 825ms/step - loss: 1.2334e-05 - acc: 1.0000 - val_loss: 1.1921e-07 - val_acc: 1.0000
Epoch 5/12
1600/1600 [==============================] - 1313s 821ms/step - loss: 3.8684e-05 - acc: 1.0000 - val_loss: 1.1921e-07 - val_acc: 1.0000
Epoch 6/12
1600/1600 [==============================] - 1311s 819ms/step - loss: 1.7143e-05 - acc: 1.0000 - val_loss: 1.1921e-07 - val_acc: 1.0000
Epoch 7/12
1600/1600 [==============================] - 1345s 841ms/step - loss: 2.3334e-05 - acc: 1.0000 - val_loss: 1.1921e-07 - val_acc: 1.0000
Epoch 8/12
1600/1600 [==============================] - 1346s 841ms/step - loss: 2.0393e-05 - acc: 1.0000 - val_loss: 1.1921e-07 - val_acc: 1.0000
Epoch 9/12
1600/1600 [==============================] - 1332s 833ms/step - loss: 1.7970e-05 - acc: 1.0000 - val_loss: 1.1921e-07 - val_acc: 1.0000
Epoch 10/12
1600/1600 [==============================] - 1322s 826ms/step - loss: 1.3359e-05 - acc: 1.0000 - val_loss: 1.1921e-07 - val_acc: 1.0000
Epoch 11/12
1600/1600 [==============================] - 1328s 830ms/step - loss: 4.3202e-05 - acc: 1.0000 - val_loss: 1.1921e-07 - val_acc: 1.0000
Epoch 12/12
 256/1600 [====>.........................] - ETA: 17:25 - loss: 2.3353e-07 - acc: 1.0000
```

**Fig. 4. Training metrics**

*The figure above is the actual output of the model after training it using the GPU provided by google. In this figure, the 'acc' refers to the training accuracy of the model, the 'val_loss' refers to the validation loss of our model per Epoch, the 'val_acc' refers to the model's prediction accuracy on the validation data sets, and the 'step-loss' refers to the efficacy of the model's prediction (the closer to 0, the better the prediction). We trained the model in 12 epochs, with 1600 samples and 400 examples for validation. The rationale behind only using 20% of our total datasets was that the model had already reached a 100% validation accuracy after being trained with 400 samples. In the first epoch, the training accuracy was the lowest at a 92% sensitivity; however, it was still able to classify the validation examples correctly. The step-loss of the model fluctuates at factor of $10^5$, after the first Epoch, which would mean that there is no significant difference in level of response of the model to training data from different cycles. This figure effectively proves the hypothesis that our CNN classifier would be able to make detections of GWs from weak signals at an accuracy higher than matched filtering*

However, it is possible that our model could have possibly been overfitted to due to the limited data available, as the training accuracy (acc) was constant (1.00) after the completion of the first training Epoch. And, in the case of our validation accuracy (val_acc), it is reasonably plausible that the accuracy remained perfect (at 1.00) throughout training, but that would signify that the model was able to clearly differentiate signal after a few hundred training samples.

Since the entirety of the model was trained on waveform templates and not actual event data, we tested our model further with actual interferometer data, to ensure that our findings translated into situations where the actual construction of GW waveforms was less uniform. We extracted the time-series for the following events from GWOSC: GW170823, GW170818, GW170817, GW170814, and GW170809. Additionally, we also uploaded 6 time-series that included a glitch (non-gaussian noise) that may have been interpreted as a GW by other methods of detection.

The model was able to classify whether there was a GW in each of the data streams correctly. Moreover, this also acts as validation that the model can determine the nuance in difference between the glitch and an actual GW. This may be partly because the waveform templates the model was trained with contained information regarding the spin, and other smaller details of the two celestial bodies.

We conducted another test to see whether the mode could use *Deep Filtering* to be able to detect a faint GW signal in a time-series contaminated by glitches that coincided in the same location as the actual wave. We merged the GW170818 event data with the highly contaminated time-series, while using different biases and weights to reposition the wave. The merged datafile was compared to the datafile that only included glitches and heavy gaussian noise. The model was able to successfully predict which of the files contained the actual GW; however, it was only one test, so we cannot generalize over all situations. These results prove the efficacy of our *Deep learning* model in making GW detection.

With regard to benchmarks in the model when using large datasets, we utilized a large portion of the O1 dataset through CernVM-FS, where the CNN's complexity enabled it to correctly identify the three events (black hole mergers) at the correctly listed times [28]. However, the sensitivity of the model reduced to 98% after large number of iterations. This further illustrates the viability for this model to be implemented into the LIGO search pipeline, in the next observing run.

## 4. CONCLUSION

The results of this study definitively support the hypothesis that this model would be able to predict the presence of GWs at a high-level of accuracy even with weak signals. This paper displays some of the latest advancements in the application of machine learning (ML) that could drastically improve the current LIGO search pipeline [17]. The model we used improves upon the previous studies done in this field, by making the process less computationally intensive whilst still increasing the sensitivity. This study is also the first to have a demonstrably higher accuracy than Matched Filtering, which is still the prevalent tool for searching for GWs. This can be inferred due to the fact that at the SNR of 4, even matched filtering cannot compete with 100% sensitivity.

In the future, LIGO would have to implement a version of a *Deep Learning* model that would monitor the real-time interferometer data and flag the presence of GWs by comparing the data of the multiple continuous data streams it has access to. Although, our model can make real-time detections of GW; there still needs to be many additions in terms of parameter estimation, and efficiency over long observing runs. This proposed parameter estimation future should be able to extract all necessary information regarding the event, in real time. However, for this to be effective, it would require a large amount of computational power.

Albeit, we made use of deep signal manifolds to train our *CNN*, it does not provide the same level of reliability as actual LIGO event data. Therefore, to improve the validity of the results in future studies, the tradeoff between the limited number of available event data, and the requirement of large amounts of training data for the model must be found. Additionally, this article demonstrates the ways in which Deep Filtering can be used to mine for a specific signal in a time-series that is contaminated by noise and glitches; thus, making this technique applicable to other unrelated fields. To expand further, this study also presents the idea that the process of fine-tuning a *CNN* could potentially be used to classify large amounts of data in real-time, assuming the model specifications are optimized.

This model is able to drastically reduce the number of false positives by including non-gaussian glitches in the training data, which equip the model to differentiate between interesting noise triggers, and real signals [29]. The use of *Deep Filtering* to improve the LIGO search pipeline is becoming increasingly important, and in the future, models would be able to conduct tasks such as glitch classification more effectively, whilst using significantly less computing power [30]. The techniques introduced in this article to classify gravitational waves using a Deep Learning may be straight forward, but their applications are endless.

## ACKNOWLEDGEMENTS

This study would not be possible without the guidance of Vivek Raghavan, who we express our utmost gratitude to for his assistance in its initiation. Furthermore, we would like to thank Haris M K from The International Centre for Theoretical Sciences (ICTS) for his insight in designing the deep signal manifolds. Additionally, we would like to make reference to the PyCBC module which was essential for the analysis of the real LIGO data. We would like to also thank Parameshwaran Ajith from ICTS for lending his support in initiating this study. Finally, we would also like to acknowledge that without the resources and insight of the LIGO Scientific Collaboration (LSC), this study would not have been able to produce the results presented in this paper.

## COMPETING INTERESTS

Author has declared that no competing interests exist.

---